\documentclass{emulateapj}
\usepackage{apjfonts}
\usepackage{graphicx}
\usepackage{amsmath}
\usepackage{amssymb}
\usepackage{amsfonts}
\usepackage{natbib}          
\shorttitle{Transient Black-Hole LMXB in Cen A}
\shortauthors{SIVAKOFF ET AL.}

\slugcomment{Accepted for publication in ApJ Letters}

\begin{document}

\title{A Transient Black-Hole Low-Mass X-Ray Binary Candidate in Centaurus A}

\author{
G.~R.~Sivakoff\altaffilmark{1},
R.~P.~Kraft\altaffilmark{2},
A.~Jord\'{a}n\altaffilmark{2,3},
A.~M.~Juett\altaffilmark{4},
D.~A.~Evans\altaffilmark{2},
W.~R.~Forman\altaffilmark{2},
M.~J.~Hardcastle\altaffilmark{5},
C.~L.~Sarazin\altaffilmark{6},
M.~Birkinshaw\altaffilmark{7,2},
N.~J.~Brassington\altaffilmark{2},
J.~H.~Croston\altaffilmark{5},
W.~E.~Harris\altaffilmark{8},
C.~Jones\altaffilmark{2},
S.~S.~Murray\altaffilmark{2},
S.~Raychaudhury\altaffilmark{9,2},
K.~A.~Woodley\altaffilmark{8},
D.~M.~Worrall\altaffilmark{7,2}
}
\altaffiltext{1}{%
Department of Astronomy,
The Ohio State University,
4055 McPherson Laboratory
140 W. 18th Avenue, Columbus, OH 43210-1173, USA;
sivakoff@astronomy.ohio-state.edu%
}
\altaffiltext{2}{%
Harvard-Smithsonian Center for Astrophysics,
60 Garden Street,
MS-67, Cambridge, MA 02138, USA
}
\altaffiltext{3}{%
Clay Fellow%
}
\altaffiltext{4}{%
NASA Postdoctoral Fellow,
Laboratory for X-ray Astrophysics, 
NASA Goddard Space Flight Center,
Greenbelt, MD 20771, USA
}
\altaffiltext{5}{%
School of Physics, Astronomy, and Mathematics,
University of Hertfordshire,
Hatfield AL10 9AB, UK
}
\altaffiltext{6}{%
Department of Astronomy,
University of Virginia,
P. O. Box 400325,
Charlottesville, VA 22904-4325, USA
}
\altaffiltext{7}{%
Department of Physics,
University of Bristol,
Tyndall Avenue,
Bristol BS8 ITL, UK
}
\altaffiltext{8}{%
Department of Physics and Astronomy,
McMaster University,
Hamilton, ON L8S 4M1, Canada
}
\altaffiltext{9}{%
School of Physics and Astronomy,
University of Birmingham,
Edgbaston, Birmingham B15 2TT, UK
}

\begin{abstract}

We report the discovery of a bright transient X-ray source, CXOU
J132518.2$-$430304, towards Centaurus A (Cen A) using six new {\it
Chandra X-Ray Observatory} observations in 2007 March--May. Between 2003
and 2007, its flux has increased by a factor of $>770$.
The source is likely a low-mass X-ray binary in Cen A with unabsorbed
0.3--10 keV band luminosities of $(2$--$3) \times 10^{39} {\rm \, erg
\, s}^{-1}$ and a transition from the steep-power law state to the thermal
state during our observations. CXOU J132518.2$-$430304 is the most luminous
X-ray source in an early-type galaxy with extensive timing information that
reveals transience and a spectral state transition. Combined with its
luminosity, these properties make this source one of the strongest candidates to
date for containing a stellar-mass black hole in an early-type galaxy. Unless
this outburst lasts many years, the rate of luminous transients in Cen A is
anomalously high compared to other early-type galaxies.
\end{abstract}

\section{Introduction}

Over the 40 years of X-ray astronomy, the detailed study of Galactic X-ray
binaries (XRBs) has placed strong constraints on theories of X-ray binary
evolution and accretion. X-ray variability studies have been important in
understanding disk accretion and the interplay between different emission
components, e.g., jet, corona, and disk.

All Galactic low mass X-ray binaries (LMXBs), which have companions of
$\lesssim 1 M_{\odot}$, that are confirmed BHs are also transient systems
with X-ray luminosities that vary by orders of magnitude \citep{RM2006}.
While most have outbursts ranging from weeks to months, one unusual transient,
GRS~1915$+$105 has been continuously emitting since its outburst began 15 years
ago. The transient behavior of BH (and some neutron star) LMXBs has been
attributed to a disk instability first identified in cataclysmic variables
\citep[CVs;][]{P1996}. While this theory correctly identifies which XRBs are
likely transients, it has yet to accurately predict the outburst lengths and
recurrence times, which are necessary to determine an X-ray duty cycle
\citep[for a review, see][]{K2006}.

The Galactic population of BH LMXBs has been studied in great detail, but the
small sample size, $\sim$190 Galactic LMXBs of which $\sim10$--25\% are
confirmed or candidate BH systems \citep{LPH2006}, and non-uniform observational
sampling limit the ability of Galactic studies to measure the duty cycle. With
the {\em Chandra X-ray Observatory}, we can now study the tens to hundreds of
identified LMXBs in individual nearby galaxies
\citep[e.g.,][]{SIB2000,SJS+2007}. With multi-epoch observations of early-type
galaxies, we can measure the variability of extragalactic LMXBs. Interestingly,
the most luminous extragalactic LMXBs are persistent on timescales of several
years \citep{Ir2006}.

Centaurus A (NGC~5128, Cen A), at $3.7{\rm \, Mpc}$ \citep[the average of 5 distance
indicators, see \S~6 in][]{FMS+2007}, is the nearest radio galaxy and the nearest optically
luminous \citep[$M_B = -21.1$;][]{DHM+1979} early-type galaxy. Prior to 2007,
{\it Chandra} had targeted Cen A with the ACIS detectors four times, 1999
December 5, 2000 May 17, 2002 September 3, and 2003 September 14 (Observations
\dataset[ADS/Sa.CXO#obs/00316]{316},
\dataset[ADS/Sa.CXO#obs/00962]{962},
\dataset[ADS/Sa.CXO#obs/02978]{2978},
and
\dataset[ADS/Sa.CXO#obs/03965]{3965}).
In 2007, {\it Chandra} performed six further deep observations ($\sim 100 {\rm
\, ks}$ each) of Cen A to study the X-ray properties of its jet, lobes, and XRBs
\citep{HKS+2007,JSM+2007,WBK+2007,KHS+2008s}.
As the nearest
luminous early-type galaxy, Cen A is one of the premier targets for studying
extragalactic LMXB variability. In this analysis, we report results on the most
luminous XRB candidate in Cen A, CXOU J132518.2$-$430304.
The errors on our spectral fits refer to
single-parameter $ 90\% $ confidence intervals and all other errors refer to $1
\sigma $ confidence intervals. All fluxes and luminosities are in the 
0.3--10 keV band.

\section{Observations and Data Reduction}
\label{sec:obs}

\begin{deluxetable}{llcccc}
\tabletypesize{\footnotesize}
\tablewidth{0pt}
\tablecaption{Chandra Observations of CXOU J132518.2$-$430304 in Cen A\label{tab:obs}}
\tablehead{
\colhead{Obs.} &
\colhead{Date} &
\colhead{Exposure} &
\colhead{OAA} &
\colhead{Net Counts} &
\colhead{Net Rate}\\
\colhead{(\#)}&
\colhead{(2007)} &
\colhead{(${\rm s}$)} &
\colhead{($\arcmin$)} &
\colhead{(${\rm cnt}$)} &
\colhead{($10^{-2} {\rm \, cnt \, s}^{-1}$)}
}
\startdata
\dataset[ADS/Sa.CXO#obs/07797]{7797} & March 22 & 96888 & 0.9 & 6471 & $6.679\pm0.084$ \\
\dataset[ADS/Sa.CXO#obs/07798]{7798} & March 27 & 90839 & 7.3 & 7942 & $8.742\pm0.100$ \\
\dataset[ADS/Sa.CXO#obs/07799]{7799} & March 30 & 94783 & 7.3 & 6981 & $7.365\pm0.090$ \\
\dataset[ADS/Sa.CXO#obs/07800]{7800} & April 17 & 90843 & 7.4 & 6259 & $6.890\pm0.089$ \\
\dataset[ADS/Sa.CXO#obs/08489]{8489} & May 8    & 93936 & 3.3 & 8987 & $9.567\pm0.102$ \\
\dataset[ADS/Sa.CXO#obs/08490]{8490} & May 30   & 94425 & 0.5 & 7908 & $8.375\pm0.095$
\enddata
\tablecomments{Observed counts are in the 0.5--$7 {\rm \, keV}$ band.
The source in observations 7797, 8489, and 8490 is at small off-axis angles
(OAAs), and their rates are underestimated due to event pileup.}
\end{deluxetable}
\begin{figure*}
\center
\includegraphics[height=170pt]{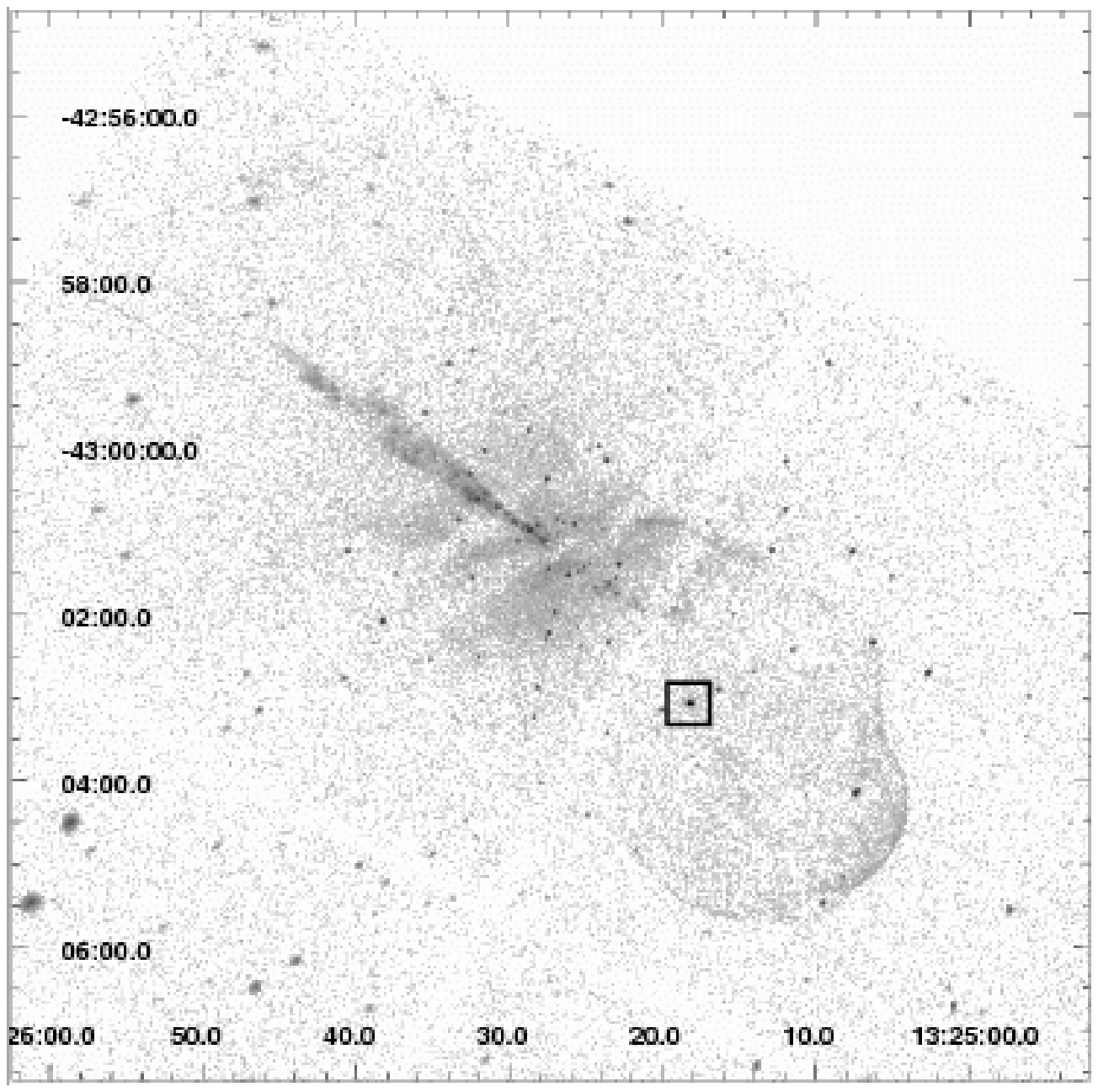}\includegraphics[height=170pt]{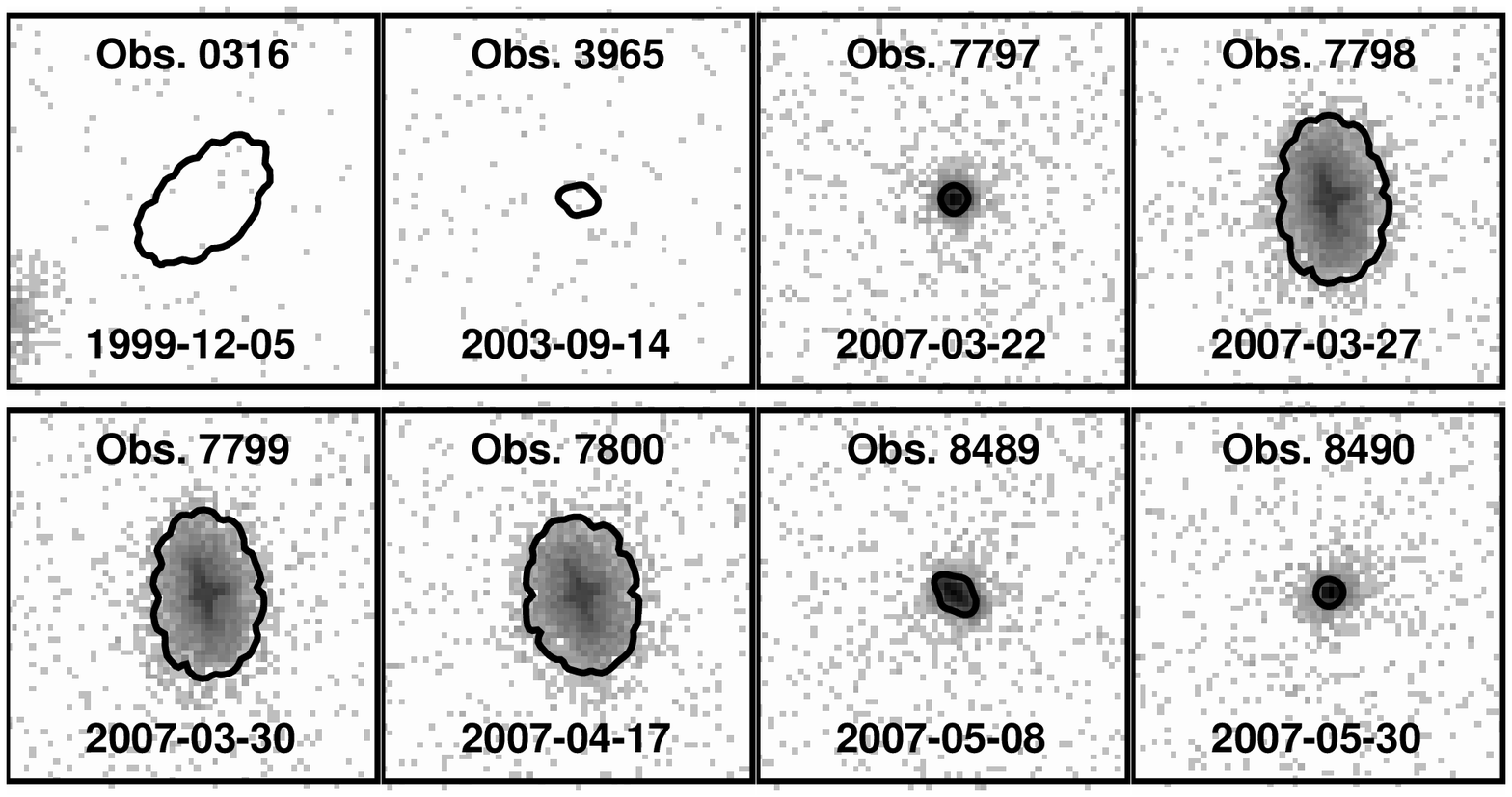}
\caption{
{\it Left:} {\it Chandra} 0.3--2 keV image of Cen A from Observation 7797, with
a logarithmic intensity scale and Gaussian smoothing (FWHM=3 pixels).
The image is limited to soft energies
to minimize the readout streak from the central AGN.
The $\sim 30\arcsec \times 30\arcsec$ box indicates
the location of CXOU J132518.2$-$430304.
{\it Right:}
{\it Chandra} 0.3--10 keV subimages
($30\arcsec \times 30\arcsec$)
of CXOU J132518.2$-$430304 with the source extraction
region indicated for all {\it Chandra} observations except
0962 and 2978, both of which show no source.
The logarithmic intensity scale
has been changed.
The source clearly went into outburst between
Observations 3965 and 7797 and has remained bright for at least
70 days. The source regions due to the varying PSFs are overlaid.
\label{fig:image}} 
\end{figure*}

New {\it Chandra} observations of Cen~A were taken
in 2007 (Table~\ref{tab:obs}) and contained no high-background periods.
Our analysis includes only events with ASCA grades of 0, 2, 3, 4, and 6. Photon
energies were determined using the gain file acisD2000-01-29gain\_ctiN0001.fits,
correcting for time dependence of the gain, charge-transfer inefficiency, and
quantum efficiency degradation. We excluded bad pixels, bad columns, and columns
adjacent to bad columns or chip node boundaries.
The absolute astrometry is accurate to $\pm0\farcs1$ \citep{HKS+2007}.
We used {\sc ciao 3.4}%
\footnote{See \url{http://asc.harvard.edu/ciao/}.}
with {\sc caldb 3.3.0.1} and NASA HEASARC's {\sc ftools 6.2}%
\footnote{See
\url{http://heasarc.gsfc.nasa.gov/docs/software/lheasoft/}%
\label{ftn:heasoft}.} for data reduction and analysis.
We used ACIS Extract 3.131%
\footnote{See \url{http://www.astro.psu.edu/xray/docs/TARA/%
ae\_users\_guide.html}} to refine the position of CXOU J132518.2$-$430304
by correlating the 0.5--$7 {\rm \, keV}$ photons near the {\sc wavdetect}
coordinates against the count weighted, combined X-ray PSF (at 1.497 keV) of the
observations. For each observation
we created source regions corresponding to a polygon
encircling 90\% of the X-ray PSF and
local circular background regions that began just beyond the region encircling
97\% of the X-ray PSF and had three times the source area. These regions were
used to extract spectra and response files for
spectral fitting using {\sc xspec 11.3.}\footnotemark[\ref{ftn:heasoft}]

\section{Properties of CXOU J132518.2$-$430304}

CXOU J132518.2$-$430304 is located $2\farcm6$ SW of the AGN in Cen A. We
estimate that its position of R.A.\ $= 13^{\rm h} 25^{\rm m} 18\fs24$ and Dec.\
$= -43\arcdeg03\arcmin04\farcs5$ (J2000) is accurate to $0\farcs2$.

In each observation, CXOU J132518.2$-$430304 was the brightest non-nuclear
source.
At this position, there is no source in any
of the four previous observations of Cen A (see Figure~\ref{fig:image}).
In those four observations, we find a combined $3\sigma$ upper limit of 14.9 net
counts. The spectral models fitted to the new observations (Table~{\ref{tab:spec}}) give
PSF-corrected, absorbed fluxes of $(8.5$--$11.5) \times 10^{-13} {\rm \, erg \,
cm}^{-2} {\rm \, s}^{-1}$. The same models correspond to fluxes of $\lesssim 1.1
\times 10^{-15} {\rm \, erg \, cm}^{-2} {\rm \, s}^{-1}$ in the old
observations. The flux has increased by a factor of $>770$.

CXOU J132518.2$-$430304 is a transient source, whose outburst began after 2003
September 14. Since it is detected in all new observations, the outburst
duration is at least 70 days.

We fitted the $0.5$--$10 {\rm \, keV}$
spectra of CXOU J132518.2$-$430304, grouping spectral channels
to have $\ge 25$
counts. Since the spectra of Galactic XRBs are often fitted with the combination
of a power-law and a multi-color disk blackbody model \citep{RM2006}, we adopted
this model attenuated by
a fixed Galactic absorption term
\citep[$N_H= 8.41 \times 10^{20} {\rm \, cm}^{2}$;][]{DL1990}
and a variable local absorption term $N_{\rm H, local}$.
The power-law photon index
$\Gamma$, the inner temperature of the disk $kT_{\rm diskbb}$, and the
normalizations of the two components were also variable parameters.
Given the high count rates, multiple
photons may be associated with an event, distorting the measured spectrum. 
Therefore, we convolved the physical model with the pileup model of
\citet{D2001}.
Its key parameters are the grade-morphing
parameter $\alpha$ and the number of independent regions $N_{\rm reg,p}$ over
which pileup is calculated. We assumed $\alpha=0.5$ for all fits, which was
consistent with fits where $\alpha$ was allowed to vary. With their off-axis
PSF, we determined that $N_{\rm reg,p}=5$ for Observations 7798--7800 and pileup
is minimal ($\lesssim 4\%$). For the nearly on-axis Observations 7797, 8489, and
8490, $N_{\rm reg,p}=1$ and pileup is a concern ($15$--$23\%$).

\begin{figure}
\center
\includegraphics[width=0.85\linewidth,clip=true]{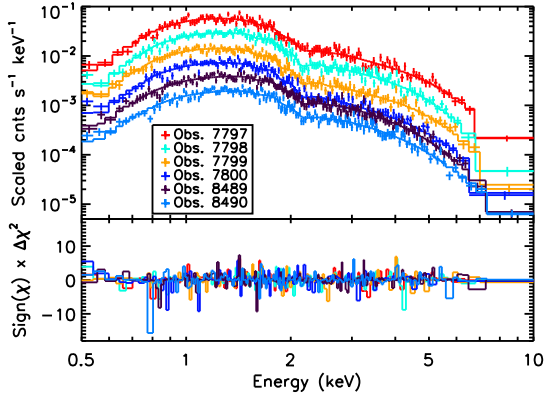}
\caption{
Spectra of CXOU J132518.2$-$430304 (0.5--$10.0 {\rm \, keV}$) with the best-fit
spectral model for each of the new observations. All spectra have been rescaled
for visibility. The spectral fit parameters are listed in
Table~{\ref{tab:spec}}. A spectral transition from the steep power-law state to
the thermal dominant state appears to occur between Observations 7800 and 8489.
\label{fig:spec}}
\end{figure}

We first attempted to jointly fit the spectra, requiring a fixed physical
spectral shape, but allowing the normalizations to vary between
observations. This fit is unacceptable with $\chi^2 = 1248.4$ for 999 degrees of
freedom (dof), suggesting that there is spectral variability
between the six observations. We next allowed each
observation to have its own spectral shape. We present the spectra in
Figure~{\ref{fig:spec}} and summarize those fits in Table~{\ref{tab:spec}}. 
Added together, the $\chi^2 = 992.1$ for 979 dof is an acceptable fit
that is significantly
preferred over models with only an absorbed power-law ($\chi^2 = 1239.9$ for 991 dof) or absorbed disk blackbody ($\chi^2 = 1416.1$ for 991 dof).
Using error-weighted averages, it is clear that the first four observations are
dominated, $L_{\rm unabs,pow}/L_{\rm unabs,tot} = 0.77\pm0.05$, by a power-law
component, $\Gamma = 2.29\pm0.08$, attenuated by a $N_{\rm H, local}$ of
$(2.3\pm0.5)
\times 10^{21} {\rm \, cm}^{-2}$.
The sub-dominant disk component has a temperature that is unconstrained in
observations 7797 and 7800, but that observations 7798 and 7799 find to be $k
T_{\rm diskbb} = 0.9 \pm 0.2$~keV. In the last two observations, a disk
blackbody component, $kT_{\rm diskbb} = 1.01\pm0.03 {\rm \, keV}$, dominates,
$L_{\rm unabs,pow}/L_{\rm unabs,tot} = 0.10\pm0.06$, over a poorly constrained
power-law component. In addition, the $N_{\rm H, local}$ is also lower,
$(0.5\pm0.3) \times 10^{21} {\rm \, cm}^{-2}$.
In Observation 8489, a very steep
power-law component with high absorption leads to poor constraints on
column density and $L_{\rm unabs,pow}/L_{\rm unabs,tot}$.

\begin{deluxetable*}{llllllllll}
\tabletypesize{\footnotesize}
\tablewidth{0pt}
\tablecaption{X-ray Spectral Fits to Individual Observations of
CXOU J132518.2$-$430304\label{tab:spec}}
\tablehead{
\colhead{Obs.} &
\colhead{$L_{\rm abs,tot}$} &
\colhead{$N_{H{\rm ,local}}$} &
\colhead{$L_{\rm unabs,tot}$} &
\colhead{$\Gamma$} &
\colhead{$L_{\rm unabs,pow}$} &
\colhead{$kT_{\rm diskbb}$} &
\colhead{$L_{\rm unabs,diskbb}$} &
\colhead{$L_{\rm unabs,pow}/L_{\rm unabs,tot}$} &
\colhead{$\chi^2/{\rm dof}$}\\
&
\colhead{($10^{39} {\rm \, erg \, s}^{-1}$)} &
\colhead{($10^{21} {\rm \, cm}^{-2}$)} &
\colhead{($10^{39} {\rm \, erg \, s}^{-1}$)} &
&
\colhead{($10^{39} {\rm \, erg \, s}^{-1}$)} &
\colhead{(keV)}&
\colhead{($10^{39} {\rm \, erg \, s}^{-1}$)} &
&
}
\startdata
7797 &  $1.49^{+0.05}_{-0.05}$ & $2.8^{+1.2}_{-1.3}$ & $2.76^{+1.12}_{-0.64}$ & $2.34^{+0.66}_{-0.44}$ & $2.63^{+0.57}_{-1.03}$ & $1.23 \, [{\rm unc}] $ & $0.13 [<0.68]        $ & $0.95 [>0.75]        $ & 140.2/151\\
7798 &  $1.69^{+0.06}_{-0.05}$ & $2.3^{+1.2}_{-1.0}$ & $2.91^{+1.14}_{-0.51}$ & $2.45^{+0.67}_{-0.37}$ & $2.20^{+0.79}_{-0.57}$ & $1.20^{+0.28}_{-0.31}$ & $0.71^{+0.39}_{-0.08}$ & $0.76^{+0.05}_{-0.08}$ & 159.4/168\\
7799 &  $1.50^{+0.04}_{-0.04}$ & $2.4^{+1.4}_{-0.6}$ & $3.42^{+1.65}_{-0.90}$ & $2.29^{+0.11}_{-0.08}$ & $2.60^{+0.31}_{-0.17}$ & $0.07 \, [{\rm unc}] $ & $0.81 [<2.15]        $ & $0.76 [>0.57]        $ & 142.6/156\\
7800 &  $1.39^{+0.04}_{-0.05}$ & $1.5^{+1.3}_{-1.1}$ & $1.99^{+0.59}_{-0.31}$ & $2.04^{+0.38}_{-0.57}$ & $1.50^{+0.92}_{-0.67}$ & $0.66^{+0.33}_{-0.12}$ & $0.49^{+0.36}_{-0.33}$ & $0.75^{+0.18}_{-0.26}$ & 163.1/144\\
8489 &  $1.88^{+0.02}_{-0.02}$ & $0.5^{+3.3}_{-0.3}$ & $2.20^{+3.69}_{-0.06}$ & $2.25 \, [{\rm unc}] $ & $0.05 [<3.75]        $ & $0.99^{+0.04}_{-0.06}$ & $2.15^{+0.02}_{-0.25}$ & $0.02 [<0.64]        $ & 186.4/186\\
8490 &  $1.80^{+0.00}_{-0.09}$ & $0.5^{+0.3}_{-0.3}$ & $2.04^{+0.05}_{-0.11}$ & $-3.00 [<0.80]       $ & $0.20^{+0.05}_{-0.12}$ & $1.05^{+0.04}_{-0.06}$ & $1.84^{+0.03}_{-0.12}$ & $0.10^{+0.03}_{-0.06}$ & 200.4/174
\enddata
\tablecomments{PSF-corrected luminosities in
the 0.3--10 keV band are calculated assuming a distance of $3.7 {\rm \, Mpc}$. 
Errors indicate 90\% confidence intervals. Brackets indicate parameter is
unconstrained in at least one direction.
An additional Galactic
$N_H = 8.41 \times 10^{20} {\rm \, cm}^{-2}$ is also included.}
\end{deluxetable*}

We used the Rayleigh statistic to search observations for periodic signals
with frequencies between $10^{-5} {\rm \, Hz}$ and $10^{-1} {\rm \, Hz}$,
testing every $10^{-5} {\rm \,
Hz}$. We only found periodic signals
related to known instrumental effects.
We looked for additional count-rate variability within each
observation using the Kolmogorov-Smirnov (K-S) test. Initial tests on the
0.3--$10.0 {\rm \, keV}$ band indicated 99.96\% significant variability in
Observation 7797 and 7800. The total flux variations are of order 10\% and 20\%,
respectively. Additional K-S tests on the 0.3--$1.0 {\rm \, keV}$ (soft),
1.0--$2.0 {\rm \, keV}$, and 2.0--$10.0 {\rm \, keV}$ (hard) bands show that the
variability in Observation 7797 (a $\sim 5 \times 10^{3} {\rm \, s}$ flare and a
lower rate over $\sim 5 \times 10^{4} {\rm \, s}$) and 
7800 (a $\sim 8 \times 10^{3} {\rm \, s}$ dip and a
higher rate over $\sim 5 \times 10^{4} {\rm \, s}$)
primarily originates in the hard and soft bands, respectively.

A {\it Hubble Advanced Camera for Surveys} observation
\dataset[ads/Sa.HST#J8Z012010]{(J8Z012010)}
 indicates that any counterpart has $m_{f606w} > 24.9 {\rm \, (AB)}$. From
{\it  Hubble WFPC2} observations
\dataset[ads/Sa.HST#U3LBA101M,ads/Sa.HST#U3LBA102M,ads/Sa.HST#U3LBA103M,
ads/Sa.HST#U3LBA104M,ads/Sa.HST#U3LBA105M,ads/Sa.HST#U3LBA106M]{(U3LBA101M -- U3LBA106M)},
the de-reddened galaxy color $(V-I)_0$ at the position of CXOU
J132518.2$-$430304 is $1.13\pm0.04$, which is consistent with K giants
\citep{P1998}. Since O/B stars would significantly alter the measured $(V-I)$
color, we rule out their presence and the possibility that CXOU
J132518.2$-$430304 is a high-mass X-ray binary. A 2007 June Very Large Array
observation places a $3\sigma$, 8.4 GHz upper limit of $0.24 {\rm \, mJy}$
(Goodger et al.\, in preparation).

\section{Discussion}

CXOU J132518.2$-$430304 is a transient source with an outburst duration of $> 70
{\rm \, days}$. Among transient sources, its large X-ray to optical flux ratio
of $\log(F_X/F_{\rm opt}) \gtrsim 3.5$ is characteristic of only strongly
absorbed AGN, CVs, and XRBs. The local column density of $N_{\rm H,
local} \sim 2 \times 10^{21} {\rm \, cm}^{-2}$ is inconsistent with the source
being a strongly absorbed AGN. If the source is Galactic in nature, its X-ray
luminosity is $\lesssim 4 \times 10^{35} {\rm \, erg \, s}^{-1}$ and any
potential companion must be a K dwarf or later in the outskirts of the Galaxy. 
Since the spectra are poorly fit by thermal bremsstrahlung models, we rule out
that CXOU J132518.2$-$430304 is a luminous Galactic CV. While the source could
be a Galactic LMXB, the field density of such sources is extremely low and
transient LMXBs are typically more luminous in outburst. Thus, it is unlikely
that CXOU J132518.2$-$430304 is a Galactic LMXB. Thus, we conclude that CXOU
J132518.2$-$430304 is most likely an XRB in Cen A. Our measured color of the
galaxy at the position of this source rules out the possibility of it being a
high-mass XRB.

Assuming CXOU J132518.2$-$430304 is a transient LMXB in Cen A, its 0.3--10.0 keV
unabsorbed luminosities are $(2$--$3) \times 10^{39} {\rm \, erg \,
s}^{-1}$, which is above the lower limit used to define an ultraluminous X-ray
source in the {\it Chandra} band \citep{IBA2004}. Although a handful of Galactic
LMXBs with BHs also have similar peak luminosities \citep{JN2004}, CXOU
J132518.2$-$430304 appears to be one of the most luminous extragalactic
transient LMXBs discovered to date. The proximity of Cen A allows us to study
this LMXB in greater detail than in any other early-type galaxy.

Since its average outburst luminosity is $\sim 17$ times the Eddington
limit for a solar mass object accreting ionized hydrogen, CXOU
J132518.2$-$430304 is more likely to be a BH-XRB than a neutron star XRB. If we
assume that the compact object mass is less than 30 $M_\odot$, CXOU
J132518.2$-$430304 is accreting at an Eddington efficiency of $\gtrsim 0.5$. 
This source appears to be a very efficiently accreting BH-XRB. 
Active BH-XRBs are often divided into three states, a thermal state
(high/soft state), a hard state (low/hard state), and the steep power-law
state (very high state) \citep{RM2006}. The steep power-law state is
characterized by a $\Gamma > 2.4$ power-law dominating over a $\sim 1 {\rm \,
keV}$ disk blackbody component, which is well matched to the spectra of CXOU
J132518.2$-$430304 in the first four observations; typically, the power-law
component in the hard state has $1.4 < \Gamma < 2.1$. In two of these
observations, we saw some evidence for intraobservation variability. The mixture
of bands that vary points to $\sim 10^{4} {\rm \, s}$ timescale variability in
both the disk and power-law components. The last two observations
are clearly best fit by the thermal state, with disk temperatures that are
consistent with Galactic BH-LMXBs \citep[e.g.,][]{MR2005}. Thus, we conclude
that CXOU J132518.2$-$430304 has likely undergone a state transition from the
steep-power law state to the thermal dominant state.
The accretion column density also changes during the transition; however, to our
knowledge no such change has been observed in Galactic BH-LMXBs. It is unclear
if the local accretion column density is actually changing or if the power-law
model does not sufficiently capture the still unknown physics of the steep
power-law state at the lower energies observed by {\it Chandra}.

Many Galactic BH-LMXBs are relatively short-term transients with infrequent
outbursts with durations of about a month \citep{MR2005}. One notable exception
is the recurring transient GX~339$-$4, which outbursts every few years. In
\citet{KKF+2001}, a luminous transient source, CXOU J132519.9$-$430317
($L_X \sim 2 \times 10^{39} {\rm \, erg \, s} ^{-1}$),
was detected 21\arcsec\ ESE of CXOU J132518.2$-$430304
in Observation
316, but was undetected five months later. The transient source 1RXH
J132519.8$-$430312, which has been associated with CXOU J132519.9$-$430317, was
luminous over 10 days in 1995, but undetected in other ROSAT observations
spanning an 8 year timescale from 1990--1998 \citep{SDE2000}. Our deep ACS
image reveals that the counterpart found by \citet{GFS+2006} is an
aggregate of several stars consistent with being red giants in Cen A.
CXOU
J132519.9$-$430317, which appears to be a recurring transient with one outburst
of at least 10 days and a second outburst with an upper limit of 835 days
\citep{SDE2000,KKF+2001}, appears similar to GX~339$-4$. 
Some BH-XRB transients, e.g., XTE~J1550$-$564 and GRO~J1655$-$40, have
intermediate-duration outbursts lasting months, while GRS~1915$+$105 has
remained bright since its outburst approximately 15 years ago \citep{MR2005}.
In Cen A, CXOU J132518.2$-$430304 is in outburst for at least 70 days
with roughly constant luminosity, 
which could be consistent with outburst timescales of months or years.

Luminous ($L_X > 8\times 10^{38} {\rm \, erg \, s} ^{-1}$) LMXBs in other
early-type Galaxies tend to be persistent sources. In \citet{Ir2006}, none of
the 18 (15) luminous LMXBs attributed to NGC 1399 (M87) were transient sources. 
This study used two (six) epochs spanning 3.3 (5.3 yr). This places a 95\% lower
limit of 50 yr for the outburst duration of such sources. On the other hand, the
two luminous non-nuclear sources in Cen A have both been identified as
transients. Since the current data do not strongly constrain the duration of
either transient, we first assumed that the previously known transient in Cen A
undergoes $100 {\rm \, d}$ outbursts and that CXOU J132518.2$-$430304 is
undergoing a long-duration (${\rm yrs}$) outburst akin to GRS~1915$+$105. In
that case, the rate of $100 {\rm \, d}$ outburst transients in Cen A is
$0.6^{+1.5}_{-0.5} {\rm \, yr}^{-1}$.
Since both NGC 1399 and M87 have absolute $K_s$-band magnitude $\approx$1.5 mags
brighter than Cen A \citep{ TDB+2001,SCS+2006}, they each have approximately 4
times the stellar mass of Cen A. By scaling the number of $100 {\rm \, d}$
outburst transients with stellar mass and considering when the galaxies were
observed, we find the rates are consistent with no such transients being found
by the
\citet{Ir2006} study of NGC 1399 and M87.

On the other hand, CXOU J132518.2$-$430304 may have an outburst duration much
shorter than decades. If we assume both Cen A sources are transients with $100
{\rm \, d}$ outbursts, the rate of such transients is $1.3^{+1.7}_{-0.8} {\rm \,
yr}^{-1}$, and we would predict the \citet{Ir2006} study should have found
$8.4^{+11.1}_{-\phn5.4}$ such transients in NGC~1399 and M87. The lower limit is
only consistent with no detected transients at the 5\% level.

As a more model independent comparison of the rates of transients, we calculate
the fraction of sources with outburst durations shorter than years. If the
outburst of CXOU J132518.2$-$430304 lasts for decades, the difference in the
fractions between Cen A ($0.5\pm0.4$) and NGC 1399 and M87 ($<0.04$) are
different only at the $80.9\%$ level. However, if the outburst is much shorter,
the fraction in Cen A is $>0.40$, while the fraction in NGC 1399 and M87 is
$<0.04$; these fractions are different at the $98.5\%$ level of confidence.

Extensive X-ray observations of the bulge of M31 have revealed $\sim$60 transient
sources \citep{TPC2006,WNG+2006,VG2007} over $\sim 4.6 {\rm \, yr}$, all
with $L_X < 4\times 10^{38} {\rm \, erg \, s} ^{-1}$. Since the
$M_{K_s}$ of Cen A is about 0.3 brighter than M31, one
expects to find $4.5^{+5.9}_{-2.8}$ luminous transients in M31, if both Cen
A transients undergo $100 {\rm \, d}$ outbursts. Thus, it is not
surprising that no such luminous transients in M31 have been observed.

New observations of Cen A
are needed to determine if CXOU J132518.2$-$430304 is undergoing
an outburst of months or years and if the rate of luminous
transients in Cen A is anomalously high compared with other early-type
galaxies.

\acknowledgments
This work was supported by NASA through {\it Chandra} award
GO7-8105X from the CXO, which is operated by the SAO
under NASA contract NAS8-03060, and {\it HST} award GO-10597 from
STScI, which is operated by AURA under NASA contract NAS5-26555.

{\it Facilities:} \facility{CXO (ACIS)}, \facility{HST (ACS/WFC,WFPC2)}, \facility{VLA}


\bibliography{ms}

\end{document}